\documentstyle[aps,preprint]{revtex}
\def\today{\ifcase\month\or
           January\or February\or March\or April\or May\or June\or
           July\or August\or September\or October\or November\or
           December\fi
           \space\number\day, \number\year}
\begin{document}
\draft
\preprint{\vbox{\hfill DOE/ER/40762--046\\
                \null\hfill UMPP \#94--033}}
\title
{Possible Tests for $b \rightarrow s g$ Penguins via
Inclusive $K$ Distributions and Exclusive Processes.}
\author{J. Milana}
\address
{Department of Physics, University of Maryland\\
College Park, Maryland 20742, USA}

\author{S. Nussinov}
\address
{Department of Physics, University of Maryland,
College Park, Maryland 20742\\
and Physics Department, Tel--Aviv University,
Ramat--Aviv, Tel--Aviv, ISRAEL}
\date{September, 1994}
\maketitle

\begin{abstract}
We discuss experimental signatures capable of nearly immediate study
that would discern/constrain new physics manifested via enhanced gluonic
penguin decays of the $b$.
\end{abstract}
\newpage
\narrowtext
\section{Introduction}

Despite the large $B$ mass, it appears that a systematic expansion in
 $\alpha_s$ and $1/m_b$ fails by about $15 \%$\cite{semilep} to predict
the semileptonic branching ratio
\begin{equation}
Br^{s.l} = \frac{\Gamma(B \rightarrow e \bar{\nu_e} X)}{\Gamma^{tot}_B}.
\end{equation}
This discrepancy presumably requires enhancement of the nonleptonic decay width
 and most likely will be resolved within the framework of the Standard Model.
For example, it could be that
local ``quark--hadron duality'' is not as good at the $B$ scale as originally
hoped.  In particular, one recent suggestion\cite{Caltech} is that
non--perturbative
effects could significantly enhance the $b \rightarrow c \bar{c} s$ rate (with
relatively slow charm quarks in the final state) beyond the naive
$\Gamma(b \rightarrow c \bar{c} s)/\Gamma(b \rightarrow c \bar{u} d) \approx
1/3$ phase space prediction.

Alternatively it has been pointed out that an enhanced $b \rightarrow s g$
rate via penguin amplitudes could help resolve the discrepancy.\cite{kagan}
These
will contribute only to the nonleptonic decay width, $\Gamma^{n.l.}$.
These decays
do not lead to charm production (and hence the reason that their contribution
could be added incoherently).  Therefore such enhanced penguin decays would not
generate an unwanted charm excess\cite{Caltech}.

A branching of $\Gamma(b \rightarrow s g)/ \Gamma^{tot}_B \approx 20 \%$ is
needed
to explain completely the semileptonic deficiency.  To leading order in
$\alpha_s$,
the gluonic penguin decay rate is given by
\begin{equation}
\Gamma (b \rightarrow s g) = \frac{8 \alpha_s}{\pi} \Gamma_o
\left( \frac{V_{bt} V^*_{ts} C_M(\mu)}{V_{bc}} \right)^2.
\end{equation}
$\Gamma_o = m_b^5 G_F^2 V_{bc}^2/ 192 \pi^3$ is the lowest order,
perturbative result for the semileptonic decay $b \rightarrow c e \bar{\nu_e}$,
(excluding phase--space modifications due to the finite mass of charm quark),
and $C_M(\mu)$ is the coefficent of the magnetic penguin operator entering the
effective Hamiltonian ($H_{eff}$)
\begin{equation}
O_g^M = {g_s\over 16\pi^2}\ m_b \bar s \tau^a G^a_{\mu \nu} \sigma^{\mu \nu}
    {1\over 2} \left (1+\gamma_5 \right) b,
\label{Mpeng}
\end{equation}
evaluated at a mass scale $\mu$ appropiate to the decay at hand.
Working within the standard model and taking
$\mu = m_b/2$, one obtains, using the evolution equations for $C_M$ found in
Ref. \cite{Ciuchini-th} (wherein the recent controversy\cite{NDR,DRED}
concerning the apparent scheme dependence of this evolution has been resolved)
that
\begin{equation}
Br(b \rightarrow s g) \approx 2 \times 10^{-3}.\label{SMinclest}
\end{equation}
We have taken $V_{ts} = .045$, the lifetime $\tau_B = 1.3$
picoseconds\cite{partdata}
and we have equated the bottom quark mass with that of the $B$ meson, {\it
i.e.},
$m_b = M_B$.  Using $m_b = M_B - .5$ yields a rate 60 \% of the above.

A significant enhancement above standard model predictions is thus required for
gluonic penguin decays to play a role in the semileptonic decay rate issue.
Admittedly this scenario may seem unlikely in view of the recent
measurement by CLEO\cite{CLEOsgamma}
of the inclusive photo--penguin decay rate, $b \rightarrow s \gamma$, found
with a
branching rate of only $Br(b \rightarrow s \gamma) \approx 2.2 \times 10^{-4}$,
and in general agreement with Standard Model predictions\cite{Ciuchini-ph}.
Nonetheless, a preferred enhancement of the gluonic penguin by some appropriate
SUSY extension of the Standard Model relative to its electromagnetic analogue
is
conceivable.  Even an enhanced branching ratio $Br(b \rightarrow s g)$
of just a few
percent could be part
of a ``cocktail''\cite{semilep} solution for the semileptonic problem.

In the following we argue that inclusive measurements of the kaon spectrum
(presumably $K_S^o$ for experimental feasibility) near the region of
$\tilde{u} = P_K/P_{Kmax} \rightarrow 1$ would be sensitive to $b \rightarrow s
g$
rates in the range $5 \% - 20 \%$.  We also discuss exclusive decay
modes, in particular $B \rightarrow K \pi$, first indications of which
were possibly seen
last year at CLEO\cite{CLEOkpi}.  Here too enhancements would naively be
likely,
although as we will see, bound state effects could complicate such
expectations.\\

\section{Inclusive K spectrum}

For a $B$ meson at rest, the inclusive kaon momentum spectrum for kaons
emerging from
the cascade
\begin{eqnarray}
B \rightarrow &D + &X\nonumber\\
	&\hookrightarrow &K + X^\prime\label{casdec}
\end{eqnarray}
can be readily calculated from the available data.  The inclusive
momentum distributions for
$B \rightarrow D(P) + X \,$\cite{BtoDinc} and
$D \rightarrow K(q) + X^\prime \,$\cite{DtoKinc}
have been experimentally measured.  A good
overall fit to these distributions can be given by the rather simple
parametrizations
\begin{eqnarray}
\rho_{D/B}(x) &=& 6 x (1 - x)\nonumber\\
\rho_{K/D}(y) &=& \frac{6}{(1-y_{min})^3} (y - y_{min}) (1 -
y)\label{probabilities}
\end{eqnarray}
where
\begin{equation}
x \equiv P/P_{max}	\hspace{.5in} y \equiv q/q_{max}
\hspace{.5in} y_{min} \approx .14
\end{equation}
and all momentum are defined in the rest frame of the decaying heavy meson.
Note that the crucial regions of interest, $x, y \rightarrow 1$ are
particularly well
fit by Eq. (\ref{probabilities}).  For simplicity we have also normalized
each distribution to one.

The energy $E_K$ of the kaon in the $B$ decay frame is obtained from the
energy and momentum in the $D$ decay frame via a boost transformation:
\begin{equation}
M_D E_K = P^0 q^0 + |\bar{P}| |\bar{q}| z
\end{equation}
where $z$ is the cosine of the angle in the rest frame of the $D$ meson between
the
decay direction of the kaon and the boost direction of the $D$.  Defining
\begin{equation}
u = E_K/E_{Kmax},
\end{equation}
and
\begin{eqnarray}
\alpha &= \frac{M_B}{P_{max}} &= \frac{2 M_B M_D}{M_B^2 - M_D^2}\nonumber\\
\beta &= \frac{M_K}{q_{max}} &= \frac{2 M_D M_K}{M_D^2 - M_K^2}\nonumber\\
a &= \frac{P_{max} q_{max}}{M_D E_{Kmax}} &=
\frac{(M_B^2 - M_D^2)(M_D^2 - M_K^2)}{2 M_D^2 (M_B^2 + M_D^2)}
\end{eqnarray}
we have then that
\begin{equation}
u = a \left [ \sqrt{x^2 + \alpha^2 }\sqrt{y^2 + \beta^2} + x y z \right
].\label{uequation}
\end{equation}
Using the fact that the $D$ is spinless so that the angular
distribution $z$ is uniform,
one obtains that the inclusive energy distribution $\rho_{K/B}(u)$, of
the cascade kaons is
given simply by
\begin{equation}
\rho_{K/B}(u) = \int_0^1 d\,x \rho_{D/B}(x) \int_{y_{min}}^1 d\,y
\rho_{K/D}(y) \int_{-1}^1 d\,z
\delta( a \left [ \sqrt{x^2 + \alpha^2 }\sqrt{y^2 + \beta^2} + x y z \right ] -
u).
\label{cascade}
\end{equation}
Integrating over the $\delta$--function using the $z$ integral, Eq.
(\ref{cascade}) becomes
\begin{eqnarray}
\rho_{K/B}(u) = \int_0^1 d\,x \frac{\rho_{D/B}(x)}{x} \int_{y_{min}}^1
d\,y \frac{\rho_{K/D}(y)}{y}
&\Theta(a \left [ \sqrt{x^2 + \alpha^2 }\sqrt{y^2 + \beta^2} + x y
\right ] - u) \times\nonumber\\
&\Theta(u - a \left [ \sqrt{x^2 + \alpha^2 }\sqrt{y^2 + \beta^2} - x y
\right ]).\label{integrals}
\end{eqnarray}
Figure (1) displays the resulting distribution $\rho_{K/B}(\tilde{u})$
which for conformity
with the rest of the literature, \cite{BtoDinc,DtoKinc}, we have plotted not as
a function of the energy variable $u$, but in terms of the kaon's momentum,
\begin{equation}
\tilde{u} = P_K/P_{Kmax}.
\end{equation}
The two remaining integrals in Eq. (\ref{integrals}) were
performed numerically and the entire distribution was again normalized
to integrate to one.

We note that at the present
only preliminary data\cite{prelim} on the $K_S^o$ inclusive spectrum at
the $\Upsilon^{4s}$
is available.  Nevertheless this does conform to the expected overall
cascade form of Figure (1).
Our main interest here however is in the endpoint region, $u \approx
\tilde{u} \rightarrow 1$
for which much more precise data will be required and should be
available from CLEO II.
In this endpoint region one can directly show that
\begin{equation}
\lim_{u \rightarrow 1} \rho_{K/B}(u) \propto (1 - u)^4.\label{falloff}
\end{equation}
For $u = 1$, both momentum variables $x,y$ must also approach their upper
limit.
The explicit distributions $\rho_{D/B}(x)$ and $\rho_{K/B}(y)$ thus
yield two of the four powers in
Eq. (\ref{falloff}).  The remaining two powers arise from phase space
constraints generated by
the $\Theta$ functions which force $x,y \ge 1 - \eta$ if $u = 1 - \eta$
and $\eta \rightarrow 0$.
The importance of this result is that $\rho_{K/B}(u)$ is severely suppressed
near $u = 1$ compared to
the much harder function for an $s$ quark jet to fragment into a kaon.
{}From general
counting rules\cite{Gunion,Carlson} one can show from perturbative QCD
that the fragmentation function for an $s \rightarrow K^{+,0}$ behaves as
\begin{equation}
\lim_{u \rightarrow 1} D_{K^{+,0}/s}(u) \sim (1-u)^2.\label{frag}
\end{equation}

Combining Eqs. (\ref{integrals}) and (\ref{frag}), the complete
distribution, $\rho_{K/B}^{tot}(u)$,
of kaons from $B$ decays, is given by
\begin{equation}
\rho_{K/B}^{tot}(u) = (1 - \epsilon) \rho_{K/B}(u) + \epsilon
D_{K^{+,0}/s}(u)\label{rhotot}
\end{equation}
where $\epsilon$ is the $b \rightarrow s g$ branching fraction.  From
the previous discussion,
one knows that for any finite $\epsilon$, at some sufficiently large $u
\rightarrow 1$,
the $s$ quark fragmentation function must dominate.  The issue now is
to better quantify
this result with a reasonable parametrization for $D_{K^{+,0}/s}(u)$.

The new CLEO $B \rightarrow X_s + \gamma$ data\cite{CLEOKgamma}
could in principle yield $s \rightarrow K$ information in a setting
which appears to be
kinematically similar to that in $b \rightarrow s g$.
Nearly $30\%$ of all the $B \rightarrow X_s + \gamma$ inclusive decays
occur through
the $K^*$ resonance.  If we simply focus on the kaons
produced through the decay of the $K^*$, one finds that the
kaon spectrum is very hard as $u \rightarrow 1$, with a typical value of $u =
.7$
(replacing $M_D$ by $M_K^*$ in Eq. (\ref{uequation}), taking $x,y \approx 1$,
 and noting that $z \approx 0$
for the $p$--wave decay $K^* \rightarrow K \pi$).  Such a contribution would
show
up quite dramatically in $\rho_{K/B}^{tot}(\tilde{u})$ for a penguin
decay rate of order $10 \%$
 and indeed would allow ready determination of $\epsilon$ in Eq.(\ref{rhotot})
to
values as small as $\epsilon = .02$ by integrating over the total kaon yield
above
$u = .7$.   In an ideal case of infinite $B$ mass the $s$ quark from $b
\rightarrow s g$
or $b \rightarrow s \gamma$ would have identical fragmentation
independent of the identity of
the recoiling system (gluon jet or single photon).
However the use of this inclusive data for the actual $B$ mass, $M_B = 5.3$GeV,
is somewhat dubious.  For $b \rightarrow s \gamma$, the relevant total
energy of the
hadronizing system of the $s$ plus spectator $\bar{q}$ is rather low,
being roughly only
\begin{equation}
W_{X_s} = \sqrt{2 E_s ``m_{\bar{q}}{\rm ''} } \approx (5.3 GeV \times
.3 GeV)^{1/2} = 1.3 GeV,
\end{equation}
as indeed is manifest by the fact that the $K^*$ and only a few other
kaon resonances
dominate the data.  Barring precocious scaling, extraction of  accurate
information
on inclusive functions such as
$D_{K^{+,0}/s}(u)$ in such kinematics is questionable.

The kinematics in the case of $b \rightarrow s g$ is, on the other
hand, different.
Assuming the $s$ quark to recoil against an oppositely moving $g
\bar{q} = \bar{3}$ source, the
total energy available for hadronization is now $M_B = 5.3$GeV and we
expect the ``leading''
$s \rightarrow K$ fragmentation to be similar to that in an $e^{+} e^-
\rightarrow s \bar{s}$
process at these energies.  Indeed comparing even just the expected
standard model rate,
Eq. (\ref{SMinclest}), with the upper bound given last year by CLEO for
$B \rightarrow K \pi$
( $< 2.6 \times 10^{-5}$) indicates that the inclusive rate will not be
dominated by just a few
resonances.

Unfortunately $s \rightarrow K$ fragmentation contributions in
$e^+ e^-$ is charged suppressed by a factor $q_s^2/\sum q_i^2$ and are
hence difficult to
extract from the known data on inclusive kaon production\cite{Argus}.
Likewise, $s$ quark jet
production in deep inelastic $\nu$ scattering
is Cabibo suppressed by $|V_{us}|^2$ and hence not particularly useful.
We are thus forced to work by analogy, guided by general symmetry principles.

Starting first from $SU(3)$ flavor symmetry, the inclusive $\pi^+$
spectrum spectrum from
$e^+ e^-$ annihilation has been recently\cite{CandW} nicely fitted
using a primary, direct
$u \rightarrow \pi^+$ term, $D_p$, and a secondary distribution term $D_s$
\begin{equation}
D_p(z,t_0) = \frac{5}{6} (1 - z)^2 \hspace{1in} D_s(z,t_0) =
\frac{5}{6}\frac{(1 - z)^4}{z}
\end{equation}
where $t_0 \equiv ln(Q_0^2/\Lambda^2)$ reflects the general fact that
in QCD these distributions
run with $Q^2$.  The motivation of the authors of \cite{CandW} for departing
from
other, perhaps more common, forms in which $D(z)$ is given jointly as one
smooth
function (e.g. as in \cite{Baier} where $D(z) = (1-z)^2/4z$) is that
the physics dictating the
two end point regions is rather different.  The logarithmic rise in
total kaon number produced
for $z \rightarrow 0$ is
 driven by mesons produced out of secondary quarks formed in the
fragmentation chain of
the outgoing quark jet.  The region $z \rightarrow 1$ is dominated by
mesons made out
of the original quark in the jet.  A separation of these two phenomena
was crucial for the
study in \cite{CandW} in which (as is the case here) the authors were
particularly interested in
the $z \rightarrow 1$ regime of the fragmentation function in order to
compare with competing
processes.

Ignoring completely secondary kaon production, and assuming for the kaon
the same primary distribution $D_p(z,t)$ as in the pion, (evolved
down from the fitted $Q_0 = 29$GeV data, to $Q = M_B$ using the analytic
expression
provided in \cite{CandW}),  the result for $\rho_{K/B}^{tot}(\tilde{u})$,
Eq. (\ref{rhotot}), using a value for $\epsilon =.2$ is shown in Fig.
(2).  For comparison
is included the result of Fig. (1) ({\it i.e.} $\epsilon = 0$), and as
is appropriate, we have
focussed only on the end--point region $\tilde{u} \rightarrow 1$.  We
see that for this
``maximal'' value of $\epsilon$, a significant difference has developed
in the expected
kaon distributions once $\tilde{u} \ge .7$.
Accurate data at larger $\tilde{u}$ allows one to probe significantly
smaller $\epsilon$,
so that for $\epsilon = .05$ comparable differences appear at $\tilde{u} =.8$
and at
$\tilde{u} = .9$, a value as small as $\epsilon = .02$ could be discerned.

Having demonstrated what should be an experimentally testable effect, it is
likely
that these estimates are however conservative.  $SU(3)$ flavor symmetry is in
fact broken and it is likely that the $s$ quark fragmentation function is
harder
than that of a $u$ quark.  As an extreme alternative, we show the
results in Fig. (3) of using
the Heavy Quark fragmentation function of Peterson {\it et al.}\cite{Peterson}
\begin{equation}
D_{K^{+,0}/s}(z) = \frac{N}{z\left [1 - 1/z - \epsilon_Q/(1-z) \right
]^2}\label{heavyfrag}
\end{equation}
in which $N$ is determined by fixing the normalization to integrate to one and
$\epsilon_Q$ is qualitatively $m^2_q/m^2_Q$, the ratio of effective
light to heavy quark masses.
Note that the $\epsilon \rightarrow \infty$ limit smoothly matches up to
the parametrization of Baier {\it et al.}\cite{Baier} for a
light--quark fragmentation function.
In the case of charm a good fit to the data\cite{heavyfragdata} is found using
$\epsilon_C = .15$.    Since in the case of the $s$ quark the choice of
$\epsilon_s$ is more
ambiguous (if at all correct),  Fig. (3) contains plots for a few
possible $\epsilon_s$ values.
As expected, the resulting distributions yield a significantly greater
departure from the cascade scenario
 of kaon production than the one indicated by Fig. (2) where perfect
SU(3) flavor symmetry
was assumed.\\

\section{Exclusive Decays}

The observation last year\cite{CLEOKgamma} by the CLEO group of the two
body exclusive decay,
 $B \rightarrow K^* \gamma$ was the first unambiguous experimental
evidence of penguin processes.
One likewise expects that rare two--body hadronic decays of the $B$--meson to
be a
particularly useful means of measuring gluon mediated, $b \rightarrow
s$ transitions.  For
concreteness, we will focus on the mode $B \rightarrow K \pi$ which, as
mentioned earlier,
was likely seen last year at CLEO\cite{CLEOkpi} (the
ambiguity involves insufficient experimental resolution
to separate candidate $\pi \pi$ from $K \pi$ decay channels).

For the two body hadronic decays, penguin processes compete with
another rare decay,
$b \rightarrow u$, which occurs at tree--level in the Standard Model
but is proportional to
 the CKM matrix elements $|V_{ub} V_{us}|^2$.
Judicious comparison with analogous processes proportional to $|V_{ub}
V_{ud}|^2$ allows
one to infer the relative importance of the gluonic penguin
contribution.  In the case
of $B \rightarrow K \pi$ the corresponding decay mode is $B \rightarrow \pi
\pi$.
(Hence the further importance that CLEO resolve these two modes, the sum of
which were reported with a total branching rate of
$2.4 \pm ^{.8}_{.7} \pm .2 \times 10^{-5}$\cite{CLEOkpi})
An observed ratio of branching rates much above (or below) the
naive $|V_{us}/V_{ud}|^2 \approx 1/20$ must be due to penguins.

Using perturbative QCD methods recently seen\cite{systematics} to give
a good description
of the two body hadronic decays of the $B$ meson, we estimate in the
standard model that
\begin{equation}
Br(B \rightarrow K \pi) \approx .5 \times 10^{-5},\label{pQCDest}
\end{equation}
not far from the CLEO data of last year\cite{CLEOkpi}.  Such a result clearly
does not allow much room for enhancement.  However, the importance of
bound--state effects
must be emphasized.  We will therefore sketch how our estimate Eq.
(\ref{pQCDest}) was
obtained, in which some particularly simplifying approximations were
used.  A full analysis
of the decay is the subject of a forthcoming work.\cite{tobedone}

Since the two body exclusive decays of the $B$ involve large mometum
transfers, they are short distance events.   A twist expansion in perturbative
QCD
suggests\cite{BrodLep} that only the contribution from the lowest order
Fock component expansions of the $B$ and of the outgoing mesons are relevant.
The
decay rate of the $B$ then involves a perturbatively calculable hard
amplitude convoluted
with a soft physics wave function, $\psi_m$, from each of the mesons.
These wavefunctions,
although as yet uncalculable from first principles, are universal for
each meson, {\it i.e.}
they factorize from the hard amplitude and hence are independent of the
process involved.
Thus as was employed in Ref. \cite{systematics}, ideally one can
phenomenologically
parametrize these wavefunctions using a (few) measured
cross--sections/decay rates. For
simplicity, we use the factorization scheme advocated by Brodsky and
Lepage\cite{BrodLep}
and take the momenta of
the quarks as some fraction $x$ of the total momentum of the parent
meson weighted by a soft physics distribution amplitude $\phi(x)$
($\phi(x)$ being then simply the quark's wavefunction $\psi(x,k_\perp)$
integrated over transverse momentum, $k_\perp$).

An important ingredient in the perturbative QCD approach to the two
body exclusive decays
of the $B$ is that the decay amplitude can acquire an
imaginary part because some heavy quark
propagators in various Feynman graphs can go on--shell in the
integration over the mesonic distribution amplitudes.  Here as
elsewhere \cite{GlennySterman}, \cite{Kahler}, picking up such poles is
legitimate in pQCD as they are not pinched singularites and hence by
the Coleman--Norton theorem \cite{CNorton}, are not associated with a
long distance event.  These poles arise (in part) because of the
factorization scheme we employ and because of what one believes to be the
correct
relation between quark and meson masses:
\begin{equation}
M_B = m_b + \Lambda,
\end{equation}
where $\Lambda \sim 500$MeV.
Since it has been found in practice\cite{systematics} that these
imaginary parts tend to
dominate the decay amplitude when they occur, we will here focus on
them to obtain our estimate Eq. (\ref{pQCDest}) with the more complete study
to be presented elsewhere\cite{tobedone}.

For our present purposes then, the most relevant graph contributing to the
decay
rate $B \rightarrow K \pi$ is shown in Figure (4).  The square indicates a
gluonic
penguin, the operator structure of which will be presently discussed.  The
cross
indicates the heavy--quark propagator that can go on--shell.
We note that graphs at this order in $\alpha_s$ other than that shown
have imaginary parts, but they are suppressed by factors of $\epsilon =
\Lambda/M_B$.
Unlike the case of the
inclusive cross--section considered in the previous section,
the restriction to explicit hadronic modes in the final state means that the
virtuality of the quark and gluon legs entering the penguin decay $b
\rightarrow s g$
cannot be discounted.  Hence more than merely
the simple chromo--magnetic penguin operator $O_g^M$, Eq. (\ref{Mpeng}),
contributes and in particular, it is found that chromo--electric penguins
\begin{equation}
O_g^E = {g_s\over 16\pi^2} \bar s \tau^a \gamma_\nu {1\over 2}
\left (1 - \gamma_5 \right) b \left (D_\mu G_{\mu \nu}\right )^a,
\label{Epeng}
\end{equation}
are highly relevant
(indeed they in fact dominate our estimate for the $B \rightarrow K
\pi$ decay rate).
Including then both operators $O_g^M$ and $O_g^E$, with Wilson coefficents
$C_M(\mu)$ and $C_E(\mu)$ respectively, the contribution (I) of the
diagram in Figure (1)
 to the decay amplitude is given by the expression
\begin{eqnarray}
(I) &= &\frac{8 A}{\epsilon_B} C_M (\mu) \int dx \, \frac{\phi_\pi (x) (1 - x)}
{x - 2 \epsilon_B - i\eta} \int dy \phi_k (y) y\nonumber \\
   &  &+ \frac{4 A}{\epsilon_B} C_E (\mu) \int dx \, \frac{\phi_\pi (x) (1 - x)
(1 + x - 2 \epsilon_B)}{x - 2 \epsilon_B - i\eta} \int dy \phi_k (y) y (1 - y)
\label{figures}
\end{eqnarray}
\noindent where $A$ is given by
\begin{equation}
A = \frac{8}{9} \alpha_s^2 f_B f_k f_\pi G_F U_{bt} U^*_{ts},
\end{equation}
 and a peaking approximation has been used for $\phi_B(z)$
\begin{equation}
\phi_B(z) = \frac{1}{2 \sqrt{3}} f_{B} \delta (z -\epsilon).\label{heavydist}
\end{equation}
$f_{B}$ is the decay constant of the $B$.

Using the distribution amplitude of Chernyak and Zhitnitsky\cite{CZpi} for the
pion
and for simplicity, the asymptotic distribution amplitude\cite{asympt}
for the kaon,\\
\begin{eqnarray}
\phi_\pi (x) &=& 5 (1 - 2x)^2\nonumber\\
\phi_k (y) &=& 1
\end{eqnarray}

\noindent the imaginary piece of Eq. (\ref{figures}) becomes:
\begin{equation}
Im (I) = \frac{5 A\pi}{\epsilon_B} (\frac{2}{3} C_E(\mu) + 4 C_M(\mu))
(1 - 2\epsilon) (1 - 4\epsilon)^2.
\end{equation}

We use $Im(I)$ to obtain our estimate of the decay rate
\begin{equation}
\Gamma^{est} = \frac{(Im (I))^2}{16 \pi M_B}.
\end{equation}
As parameters we use
\begin{eqnarray}
V_{ts} &=& .045 \nonumber \\
f_B &=& \sqrt{2} f_\pi \nonumber \\
\epsilon &=& \frac{.5}{M_B} \nonumber\\
\Lambda_{QCD} &=& .2 GeV.
\end{eqnarray}
For the scale $\mu$ we take the virtuality of the softest gluon exchanged,
$\mu^2 \approx .5(GeV)^2$.
For the evolution of $C_E(\mu)$, only Cella {\it et al.} of Ref. \cite{NDR}
have calculated the anomalous dimension mixing matrix relevant for the
chromo--electric penguins, and thus we use their results.  This might be
thought imprudent, considering the
controversy that was associated with the chromomagnetic
penguins\cite{Ciuchini-th,NDR,DRED}.
We note however that in the case of the chromomagnetic penguins,
the final results of Ciuchini {\it et al.}\cite{Ciuchini-th}
differ only slightly from that of Cella {\it et al.} for those
operators mutually calculated,
and hence one might expect that any errors due to scheme dependence
would be kept at a minimum.   Such is our hopes in the present work, although
we acknowledge and stress the importance that these expectations be confirmed.

With these inputs we obtain that
\begin{equation}
Br^{I} (B \rightarrow K \pi) \approx .3  \times 10^{-5},
\end{equation}
where the superscript $I$ reminds the reader this estimate is based
only on picking up
the (leading) imaginary part of the decay amplitude.  Roughly assuming that
the real part (which adds incoherently but involves many more graphs
and also in principle
 more off--shell operators) is of the same order, we obtain Eq.
(\ref{pQCDest}).

We note that similar such bound--state effects
have been previously considered\cite{pengleav} in the case of
$B \rightarrow K^* \gamma$ decays,
however there the physical on--shell photon eliminates the electromagnetic
version of Eq. (\ref{Epeng}).  Only bound--state effects involving the
off--shell
character of the quark propagators were thus needed to be considered.
Although large in amplitude, these amusingly were found, due to accidents of
phase,
to have minimal effect on the decay rate assuming standard model parameters
of $H_{eff}$.
The purely hadronic two body decay $B \rightarrow K \pi$
does not however enjoy such simplifying features.  Chromoelectric penguins are
found to be highly relevant because the restriction to a $K \pi$ final
state allows for the gluons to be
significantly off--shell and because the Wilson coefficient of the
chromo--electric
operator is appreciablely larger than that of the chromo--magnetic
operator (at either the $M_W$ scale or when ran to lower scales
appropriate for the decay of the $B$).

The complications of bound--state
effects and in particular the introduction of additional operators thus allows,
at least in principle, significant differences between the
inclusive rates discussed in section (II) and that of any particular decay mode
as
discussed here.  One tenable although perhaps contrived scenario is that
{\it only} $O_g^M$ is significantly enhanced.
Although somewhat bizarre, especially
 since for the gluon penguins it is at a fundamental level the
same Feynman graphs that determine
the Wilson coefficients of both the chromomagnetic as well as
chromoelectric operators,
such a scenario cannot nevertheless be precluded.
Indeed the fact that QCD corrections play an important role and
produce significant changes in both the absolute magnitudes and relative sizes
 of the various coefficients, means that such a scenario might yet be feasible.
However one should recall
that this preferential treatment would also have to extend into the
electromagnetic sector
where the data\cite{CLEOsgamma,CLEOKgamma} also does not allow
significant departure from standard model predictions.\\

\section{Conclusions}

It is well known that deviation of $b \rightarrow s \gamma$ and $b \rightarrow
s g$
amplitudes from standard model estimates could directly indicate new
physics.  The study
of potential $b \rightarrow s g$ enhancement is further motivated by its
possible
contribution to resolving the $B$ semileptonic problem.  We have here
discussed the possibility
of detecting such enhancement and estimating the overall strength of $b
\rightarrow s g$
transitions by focusing on the resulting inclusive kaon distribution in
the region near
$P_K = P_K^{max}$ and alternatively looking at the extreme case of $P_K
= P_K^{max}$
corresponding to the (penguin generated) $B \rightarrow K \pi$
exclusive final state.

Clearly future studies could make use of the much richer topologies in
an effort to have
a more sensitive extraction of $b \rightarrow s g$.
\footnote{We are indebted to H. J. Lu for this observation.}
Remarkably however, mere use of the inclusive kaon distribution will be
sufficiently
sensitive to an enhanced $b \rightarrow s g$ total rate of even just a
few percent as
the significantly harder kaons near $P_K = P_K^{max}$ that
result from the penguin decays would dominate over the softer kaons from
the ordinary cascade $B \rightarrow D (+ X) \rightarrow K (+ X^\prime)$.

The exclusive $B \rightarrow K \pi$ decay mode is sensitive to the
chromoelectric
penguin term.   Present data and standard model
estimates already would appear to exclude enhancement of this operator by more
than a factor of two or so, although potential uncertainties exist
concerning the evolution
of such ``off--shell'' operators that may yet modify these estimates.
Even barring such complications, these conclusions while suggestive
nevertheless do not preclude a
significantly large enhancement in the total, inclusive production of kaons as
the
chromoelectric operator vanishes for on--shell gluons and hence does not
contribute
to the perturbatively calculated, inclusive $b \rightarrow s g$ decay rate
(obtained using chromomagnetic penguin transitions).

Given these initial results and in view of the fact that enhanced
$b \rightarrow s g$ decays are indications of new physics,
the simple first step of looking for it via the inclusive kaon spectrum,
complemented by better data on the exclusive two--body decays, seems an
undoubtedly
worthwhile enterprise.

\bigskip
{\centerline{ACKNOWLEDGEMENTS}}
\medskip
We would like to thank H. Jawahery for many useful discussions and
H. Nelson, J. Rosner, and R. Schindler for their help.  Special thanks
also to C. E. Carlson.  This work was supported in part by DOE Grant
DOE-FG02-93ER-40762.

\begin{figure}
\caption{The expected distribution of kaons (normalized to one) with momenta
coming
from the cascade decay Eq. (\protect{\ref{casdec}}) of the $B$ meson.}
\end{figure}

\begin{figure}
\caption{The expected distribution of kaons if 20 $\%$ of the decays
arose via gluonic
penguins and the $s \rightarrow K$ fragmentation function was SU(3) symmetric
with
 $u \rightarrow \pi^+$.  In dots, the tail of Fig. (1) has also been
included for comparison.}
\end{figure}

\begin{figure}
\caption{Same as Fig. (2) except a Heavy Quark fragmentation function has been
used
for $s \rightarrow K$.  The various plots, from top to bottom, are for
$\epsilon_s = 1, 4, 10$ respectively, in Eq. (\protect{\ref{heavyfrag}}).}
\end{figure}

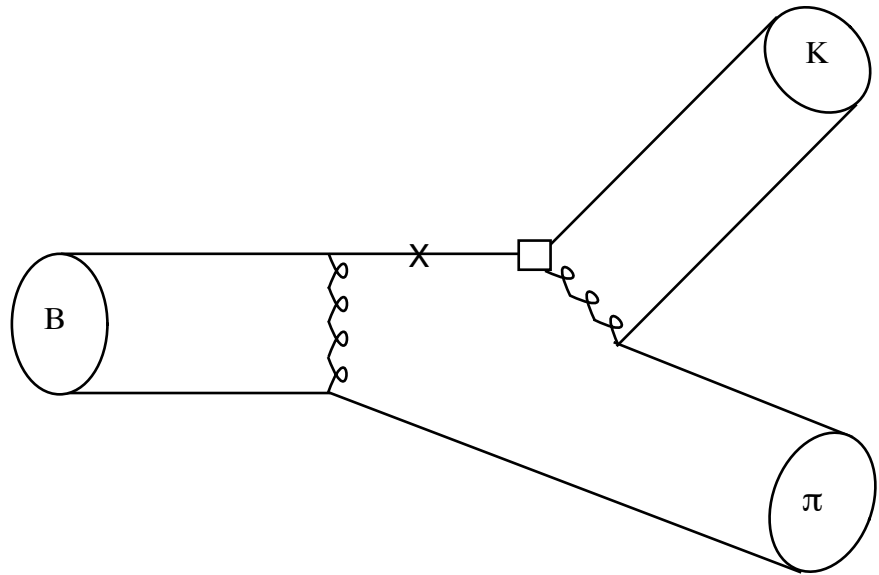
\begin{figure}
\caption{The diagram with largest imaginary phase contributing to $B
\rightarrow K \pi$
in a perturbative QCD analysis.  The cross indicates the heavy quark propagator
that can go on--shell.  The square represents gluonic penguin operators
from $H_{eff}$.}
\end{figure}
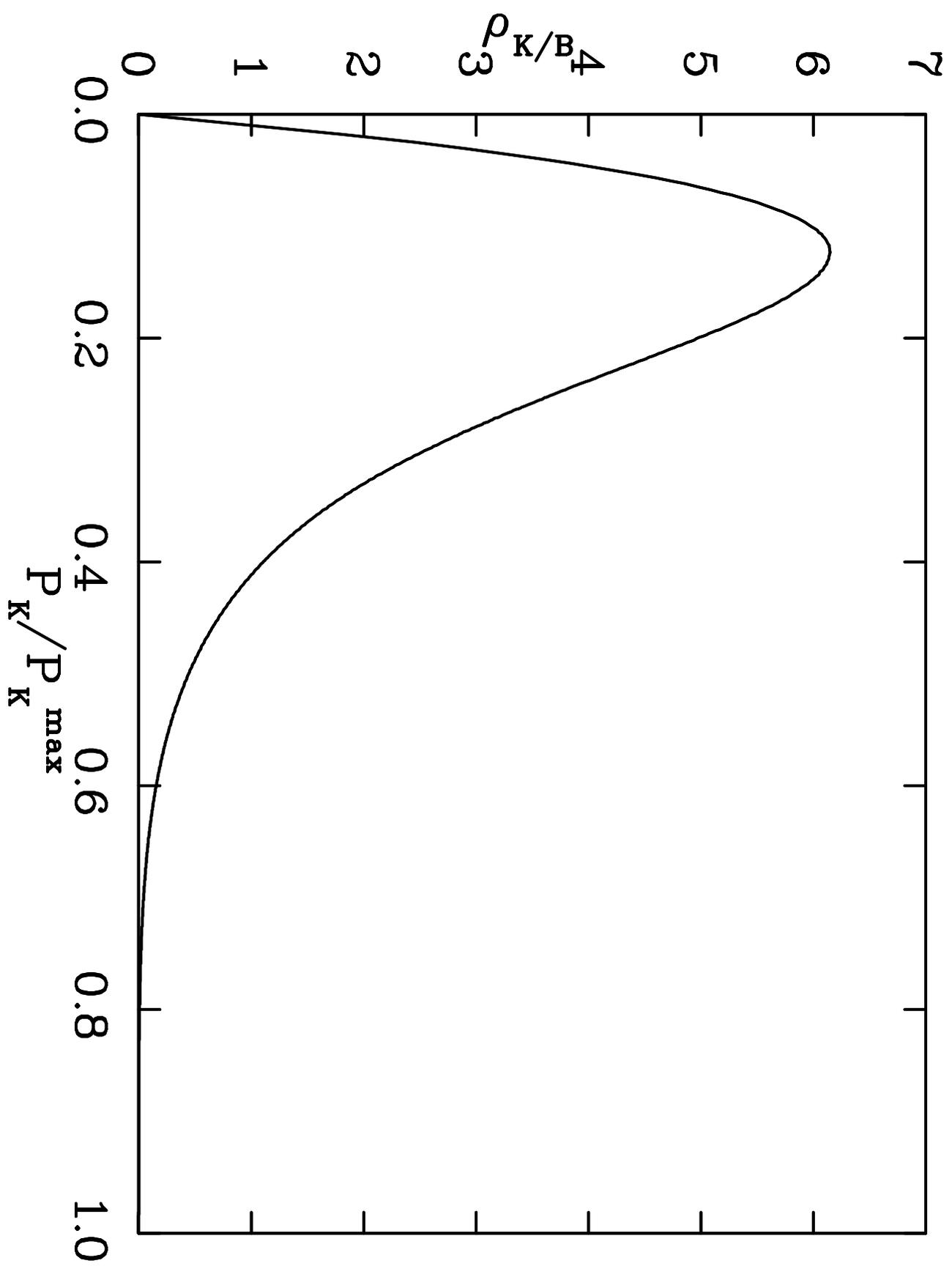
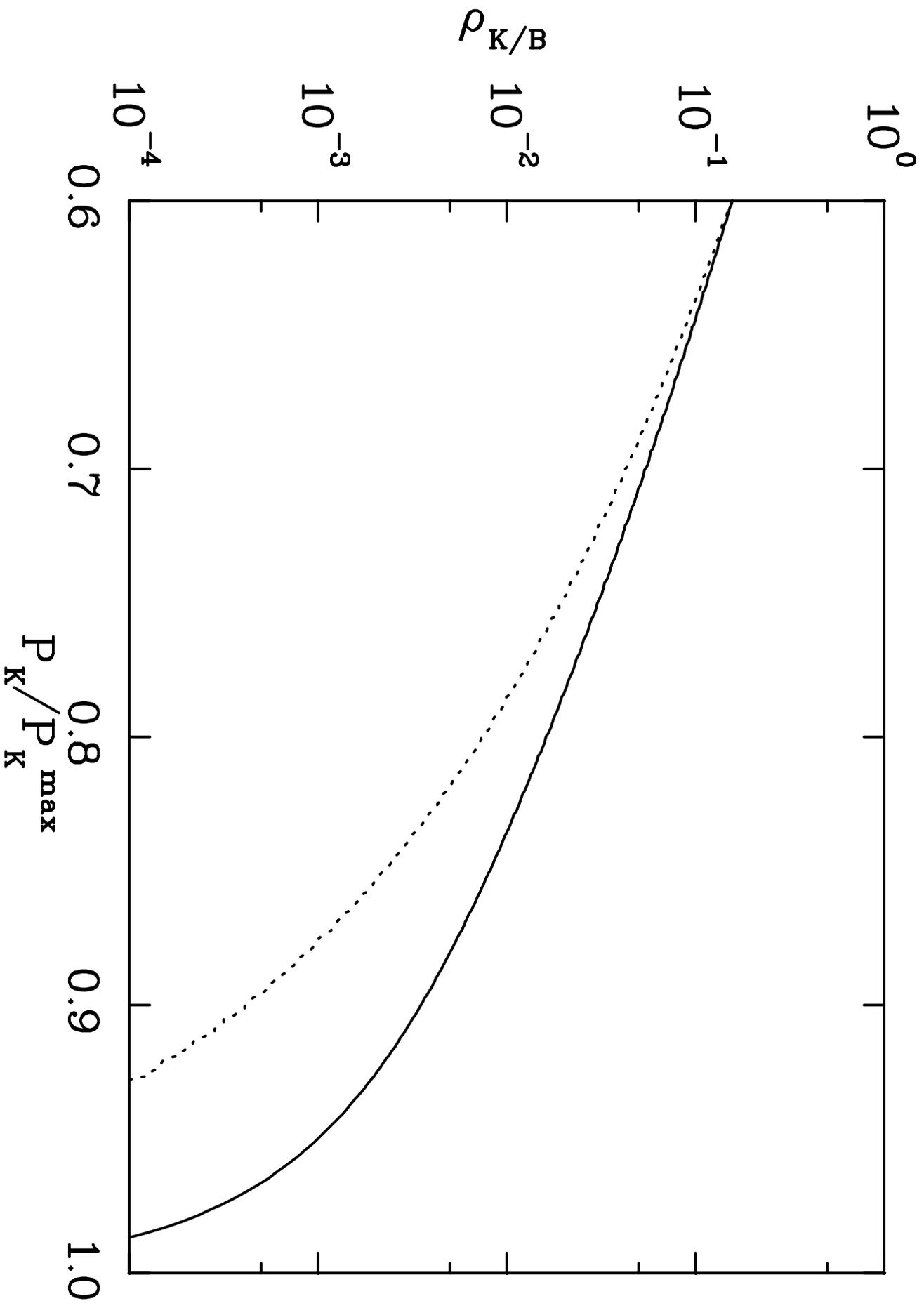
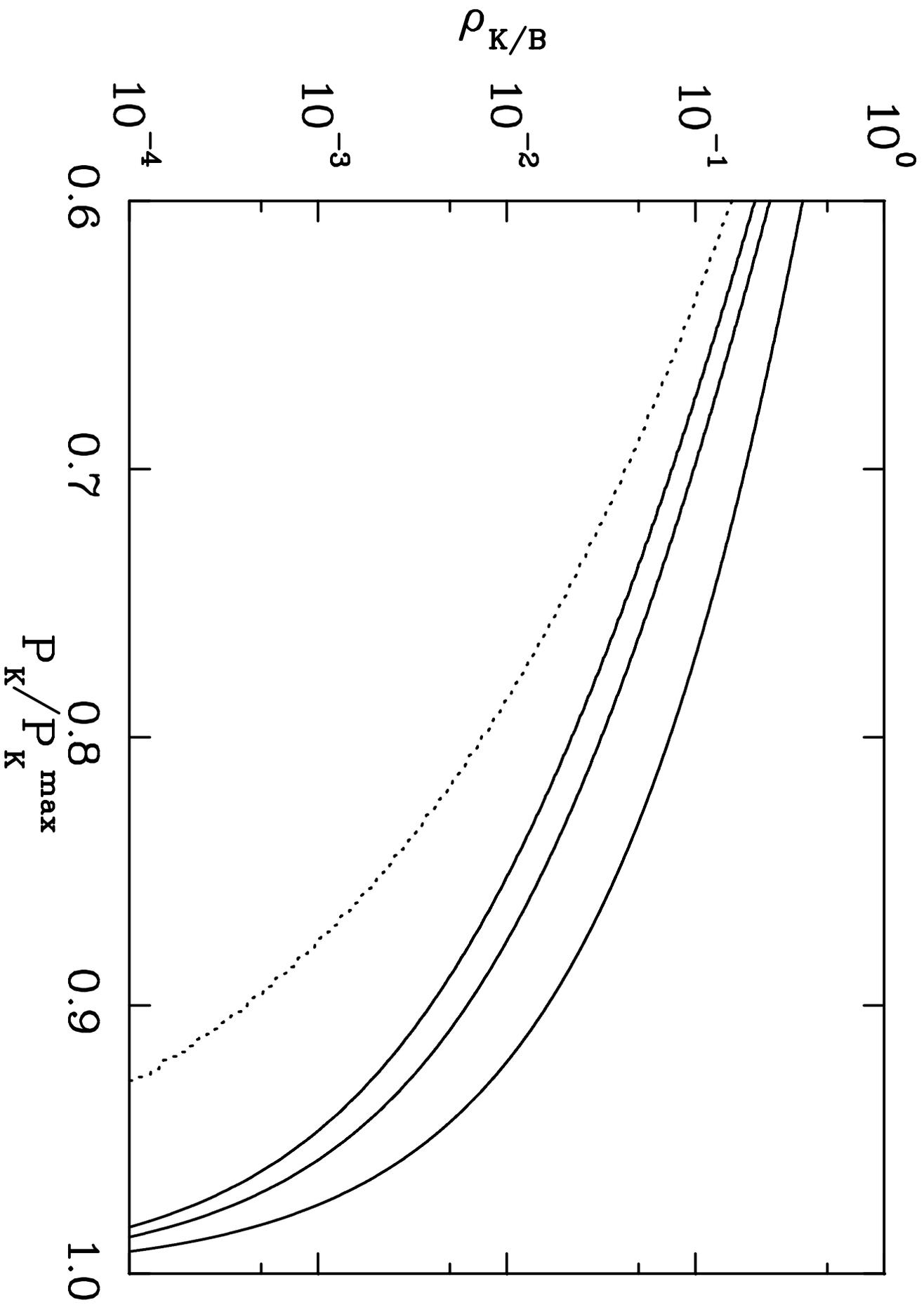
\end{document}